\documentclass[prd,twocolumn,a4paper,superscriptaddress,floatfix]{revtex4}
\usepackage{graphicx}
\begin{document}

\newcommand{\be}{\begin{equation}}
\newcommand{\ee}{\end{equation}}
\newcommand{\bq}{\begin{eqnarray}}
\newcommand{\eq}{\end{eqnarray}}
\newcommand{\bsq}{\begin{subequations}}
\newcommand{\esq}{\end{subequations}}
\newcommand{\bc}{\begin{center}}
\newcommand{\ec}{\end{center}}

\title{Effects of Inflation on a Cosmic String Loop Population}

\author{P.P. Avelino}
\email[Electronic address: ]{ppavelin@fc.up.pt}
\affiliation{Centro de F\'{\i}sica do Porto, Rua do Campo Alegre 687, 4169-007 Porto, Portugal}
\affiliation{Departamento de F\'{\i}sica da Faculdade de Ci\^encias
da Universidade do Porto, Rua do Campo Alegre 687, 4169-007 Porto, Portugal}
\author{C.J.A.P. Martins}
\email[Electronic address: ]{C.J.A.P.Martins@damtp.cam.ac.uk}
\affiliation{Centro de F\'{\i}sica do Porto, Rua do Campo Alegre 687, 4169-007 Porto, Portugal}
\affiliation{Department of Applied Mathematics and Theoretical Physics, Centre for Mathematical Sciences,\\ University of Cambridge, Wilberforce Road, Cambridge CB3 0WA, United Kingdom}
\author{E.P.S. Shellard}
\email[Electronic address: ]{E.P.S.Shellard@damtp.cam.ac.uk}
\affiliation{Department of Applied Mathematics and Theoretical Physics,
Centre for Mathematical Sciences,\\ University of Cambridge,
Wilberforce Road, Cambridge CB3 0WA, United Kingdom}

\date{11 October 2007}
\begin{abstract}
We study the evolution of simple cosmic string loop solutions in an inflationary universe. We show, for the particular case of circular loops, that periodic solutions do exist in a de Sitter universe, below a critical loop radius $R_c H=1/2$. On the other hand, larger loops freeze in comoving coordinates, and we explicitly show that they can survive more $e$-foldings of inflation than point-like objects. We discuss the implications of these findings for the survival of realistic cosmic string loops during inflation, and for the general characteristics of post-inflationary cosmic string networks. We also consider the analogous solutions for domain walls, in which case the critical radius is $R_c H=2/3$.
\end{abstract}
\maketitle

\section{\label{intr}Introduction}
Topological defects are unavoidably formed at cosmological phase transitions \cite{Kibble,Book}. Studying their physical properties, evolution and cosmological consequences is therefore mandatory for a proper understanding of the early universe. The last three decades have seen dramatic progress in this task (see \cite{Book} for a review), but significant knowledge gaps still exist. The aim of the present paper is to eliminate one of these gaps.

Defect-forming phase transitions often occur near or at the end of inflation. Moreover, it is possible that various stages of inflation occur, with defects being formed in between them \cite{openinf}. It is therefore important to understand the effects of inflation on the defects, as well as to quantify their ability to survive any inflationary periods that occur after they form. The effects of inflation on the internal (microscopic) structure of defects have been studied in \cite{basu}, which shows that those with thickness $\delta H>1/\sqrt{2}$ get smeared by expansion, while those with smaller thickness survive. Effects on macroscopic (cosmological) scales are well known for the case of long string networks \cite{quantitative,extending}, but this is not so for the loop populations, despite the fact that they are known to contain a total amount of energy which is comparable to (if not greater than) that in the long strings \cite{quantitative,protyloops}.

Here we study the effects of inflation on specific loop solutions, and consider both their microscopic and macroscopic evolution. We also discuss the consequences of our findings for cosmological scenarios involving string networks. Incidentally, we note that loops in a flat anisotropic universe were studied in \cite{anisot}, where it was shown that the anisotropy of the background has an effect on the loop motion; the reasons for this will become clearer in what follows. We will also present a brief analysis of the analogous solutions for domain walls. We shall assume that the source of inflation is a perfect fluid with equation of state $p=w\rho$ (with $ w < -1/3$) and the scale factor behaves as $a \propto t^{2/3(1+w)}$; if $w = -1$ then $a \propto \exp(Ht)$ with $H$ being constant. 

\section{\label{csevol} Microscopic loop evolution}

The world history of a Goto-Nambu cosmic string \cite{Kibble} can be represented by a two-dimensional world-sheet \cite{Book} $x^\nu = x^\nu (\sigma^a)$, with $a = 0,1\, \nu = 0,1,2,3$, obeying the usual Goto-Nambu action. In a flat FRW universe the line element is given by $ds^2=a^2(\eta)(d\eta^2 - {\bf d x}^2)$, where $a$ is the scale factor, $\eta$ is the conformal time and ${\bf x}$ are conformal spatial coordinates. If we identify conformal and world-sheet times and require that the string velocity be orthogonal to the string direction (\textit{i.e.} $\dot {\bf x} \cdot {\bf x}^\prime = 0$) then the string equations of motion take the form \cite{Book,first,quantitative}
\be
\ddot {\bf x}+ 2{\cal H}\left(1-{\dot {\bf x}}^2\right)\dot {\bf x} =\frac{1}{\epsilon}\left( \frac{{\bf x}^\prime}{\epsilon}\right)^\prime\,,\dot \epsilon + 2 {\cal H}{\dot {\bf x}}^2\epsilon = 0
\label{stringt}
\ee
where the coordinate energy per unit length, $\epsilon$, is $\epsilon^2 = {\bf x}^{\prime\,2}/(1- {\dot {\bf x}}^2)$,
${\cal H}={\dot a}/a$, and dots and primes are derivatives with respect to the conformal time and space coordinates. 

Now consider the evolution of a circular cosmic string loop. Its trajectory can be described by ${\bf x}(\eta,\theta) = q(\eta) \, \left(\sin\theta, \cos\theta,0\right)$ with $0\leq \theta\leq 2 \pi$. Let us define the \textit{invariant} loop radius as $R=|q| a \gamma$, proportional to the energy of the loop (it is also useful to define the \textit{physical} radius, $r=aq$) and $\gamma=(1-v^2)^{-1/2}$, with $v={\dot q}$ being the microscopic speed of the loop. Then the microscopic string equations of motion (\ref{stringt}), written in terms of $R$, $v$ and physical time $t$, are
\be
\frac{dR}{dt}=(1-2v^2)HR\,,
\frac{dv}{dt}=(1-v^2)\left(\frac{f(v)}{R}-2Hv\right)\,
\label{micvloop}
\ee
where $H={\cal H}/a$ is the Hubble parameter and $f(v)=\gamma sign{(-q)}$.

\begin{figure}
\includegraphics[width=3.5in,keepaspectratio]{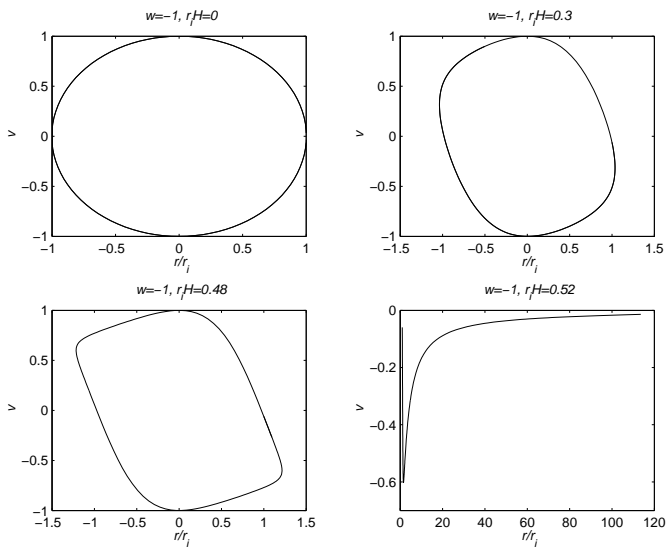}
\caption{\label{figure1}Phase space evolution of four circular loops (with different initial sizes) in an Einstein-de Sitter universe.}
\end{figure}

\begin{figure}
\includegraphics[width=3.5in,keepaspectratio]{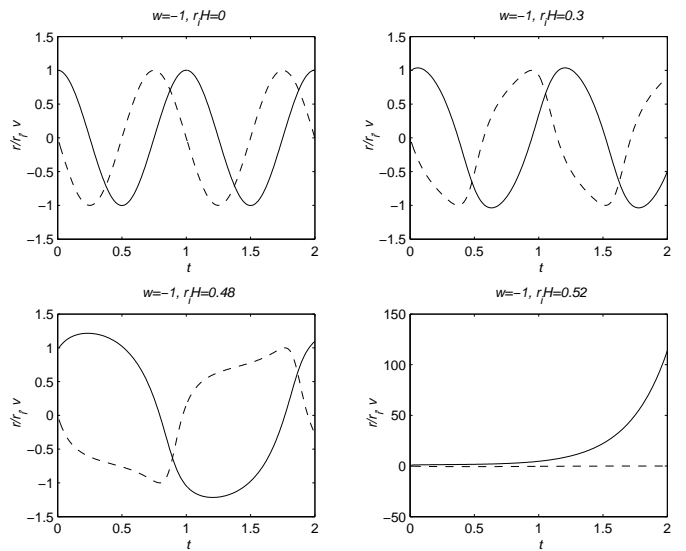}
\caption{\label{figure2}The time evolution of the loop radius and velocity,
for the loops of Fig. \protect\ref{figure1}. The solid and dashed lines
correspond to the loop size and velocity, respectively.
Time is in units of the initial loop length.}
\end{figure}

We start by studying the evolution of four different initially static circular loops (with a range of initial sizes) in a de Sitter universe ($w=-1$). In Fig. \ref{figure1} we plot their phase space diagrams $(r/r_i,v)$: their motion is periodic as long as the loop is small enough compared with the Hubble radius, that is
\be\label{halfsize}
r_i H < R_c H= \frac{1}{2}\,.
\ee
This happens because if $H$, is a constant then Eqns. (\ref{micvloop}) are invariant with respect to the transformation $v \to -v$, $f \to - f$. This symmetry of the circular loop equations of motion implies that if the loop is able to collapse, which is the case for $r_i < R_c$ (as in this regime the effect of curvature dominates over the maximum Hubble damping effect), then the loop motion will necessarily be periodic. One can therefore say that the motion of a loop with initial radius smaller than the critical radius given by (\ref{halfsize}) never becomes \textit{dominated} by the background cosmology. Nevertheless, smaller loops than this critical size do get \textit{affected} by it. This can be easily seen, for example in Fig. \ref{figure2} where we plot the loop evolution as a function of cosmic time: the distortion of the curves caused by the very rapid expansion of the universe is evident. In particular the loop period becomes larger as we increase their size, approaching infinity when $r_i \to R_c$.

For periodic solutions $\langle d(\ln R)/dt \rangle = 0$ and consequently it follows from Eqn. (\ref{micvloop}) that $\langle v^2 \rangle = 1/2$ if $H$ is a constant. Here the brackets denote a time average over one period. This is expected since the energy density in sub-critical periodic loops should evolve as matter ($\rho \propto a^{-3})$ assuming a negligibly small loop production from the long string network. Since the cosmic strings equation of state is $w \equiv p/\rho =(2\langle v^2 \rangle -1)/3$ then $\langle v^2 \rangle = 1/2$ is just what we need for $w=0$.

There is also a stationary solution of Eqns. (\ref{micvloop}) with fixed physical radius and velocity
\be
Hr = 1/\sqrt{2}\,,\qquad {\dot r} = 0\,. \label{statphys}
\ee
This may appear to be static but it's only stationary---it is a contracting loop standing still against the Hubble expansion. Note that in comoving coordinates one has ${\cal H}q = 1/\sqrt{2}$ and $v = {\dot q} = -1/\sqrt{2}$. Above we assumed the loops to be initially static ($v_i=0$), which does not include this limiting case. However, with $v_i=0$ and $r_i = R_c$, our solutions asymptote to the stationary solution. For larger loops with $r_i > R_c$ the motion is not periodic and the loop will freeze in comoving coordinates. Specifically, the loop will asymptotically behave as
\be\label{largerr}
q=const. \Longleftrightarrow R\propto a\,,\quad
v\propto a^{-1}\,.
\ee
This behavior is similar to the corresponding scaling law for the long string network, discussed in \cite{extending}.

\begin{figure}
\includegraphics[width=3.5in,keepaspectratio]{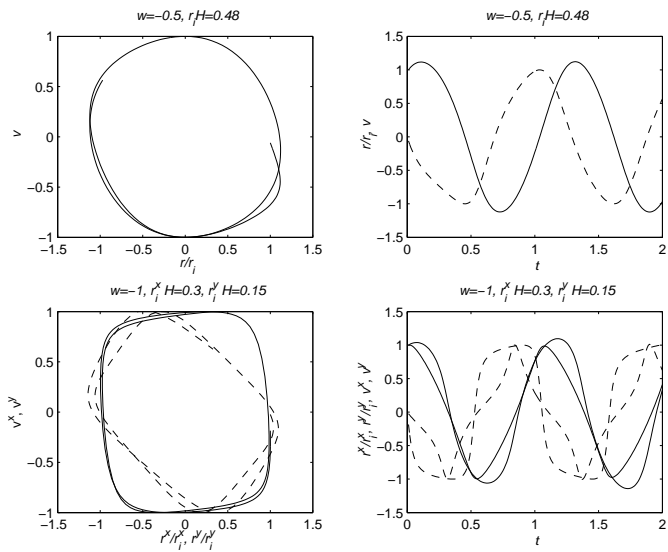}
\caption{\label{figure3}Phase space diagram and time evolution of
a circular loop with near-critical size in the case $w= -0.5$ (top) and of an initially elliptic loop
with a major to minor axis ratio of 2 (bottom). Solid and dashed lines respectively correspond
to the loop size and velocity, and time is in units of the initial loop
length. For the elliptic loop there are two sets of curves, corresponding
to projections along the major and minor axes (respectively x and y).}
\end{figure}

The situation is more complicated if $w \neq -1$ and/or for non-circular loops: then the loop motion will no longer be periodic. An example can be seen in Fig. \ref{figure3}. Given that for $w > -1$ the Hubble radius will be a growing function of time, the de Sitter case will actually be the best possible situation for loop survival in the absence of phantom behavior. In all other cases, loop survival will be more difficult and for $R<R_c=0.5 H^{-1}=0.75(1+w)t$ the loops slowly disappear by decaying into gravitational radiation. Although we only study very simple loop solutions we expect these results to approximately hold for realistic loops chopped off by a cosmic string network. Most of these are produced on scales smaller than the horizon, typically at the scale of the string correlation length if there is no significant amount of small-scale structure (which we would expect after a few e-folds), or even on much smaller scales if small-scale wiggles are present. However such very small loops will be irrelevant as far as the ability of cosmic strings to survive an inflationary era is concerned.


\section{\label{numsim}Averaged string evolution}

The averaged properties of a cosmic string network can be accurately described in the context of the velocity-dependent one-scale (VOS) model \cite{first,quantitative,extending}. It describes the string dynamics in terms of two macroscopic parameters, the string root-mean-squared (RMS) velocity, $v$, and the string correlation length, $L$, which is assumed to be the same as the string curvature radius and is related to the string energy density via $\rho=\mu/L^2$ (with $\mu$ being the string mass per unit length). We note that we will be using the variable $v$ to denote both the microscopic and the string RMS velocity depending respectively on whether the microscopic or the averaged cosmic string evolution is considered. 

In theis context the evolution equations for these quantities have the following form \cite{first,quantitative}
\be
2\frac{dL}{dt}=2HL(1+v^2)+{\tilde c}v\,,
\frac{dv}{dt}=(1-v^2)\left(\frac{k(v)}{L}-2Hv\right)\,,
\label{vosv}
\ee
where ${\tilde c}$ is the loop chopping efficiency and $k(v)$ is a function of velocity known as the momentum parameter \cite{extending}.  One can also use the model to describe the evolution of a given cosmic string loop; circular loops are characterized by their radius $R$ (now an averaged quantity). In this case the evolution equations are
\be
\frac{dR}{dt}=(1-2v^2)HR\,,
\frac{dv}{dt}=(1-v^2)\left(\frac{k(v)}{R}-2Hv\right)\,.
\label{vosvloop}
\ee
Note the obvious similarities with (\ref{micvloop}).

\begin{figure}
\includegraphics[width=3.5in,keepaspectratio]{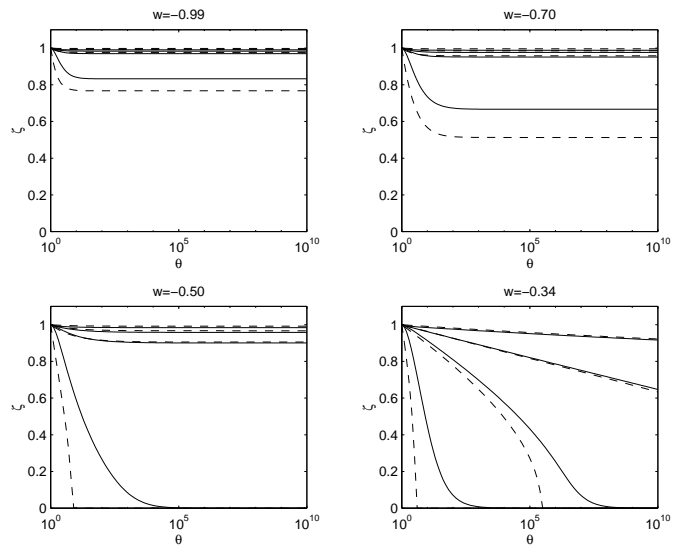}
\caption{\label{figure4}Comparing the full (numerical, Eqns.
(\protect\ref{vosvloop})) and approximate
(analytic, see Eqn. (\protect\ref{sol1loop})) VOS model solutions for the
evolution of large loops during an inflationary epoch. 
These are respectively shown by solid and dashed lines, for each set of
initial conditions. Each panel corresponds to a different value of
the equation of state parameter $w$, and within each one the four sets
of lines correspond to initial loop sizes $\epsilon=10,5,3,1$,
respectively from top to bottom.}
\end{figure}

\begin{figure}
\includegraphics[width=3.5in,keepaspectratio]{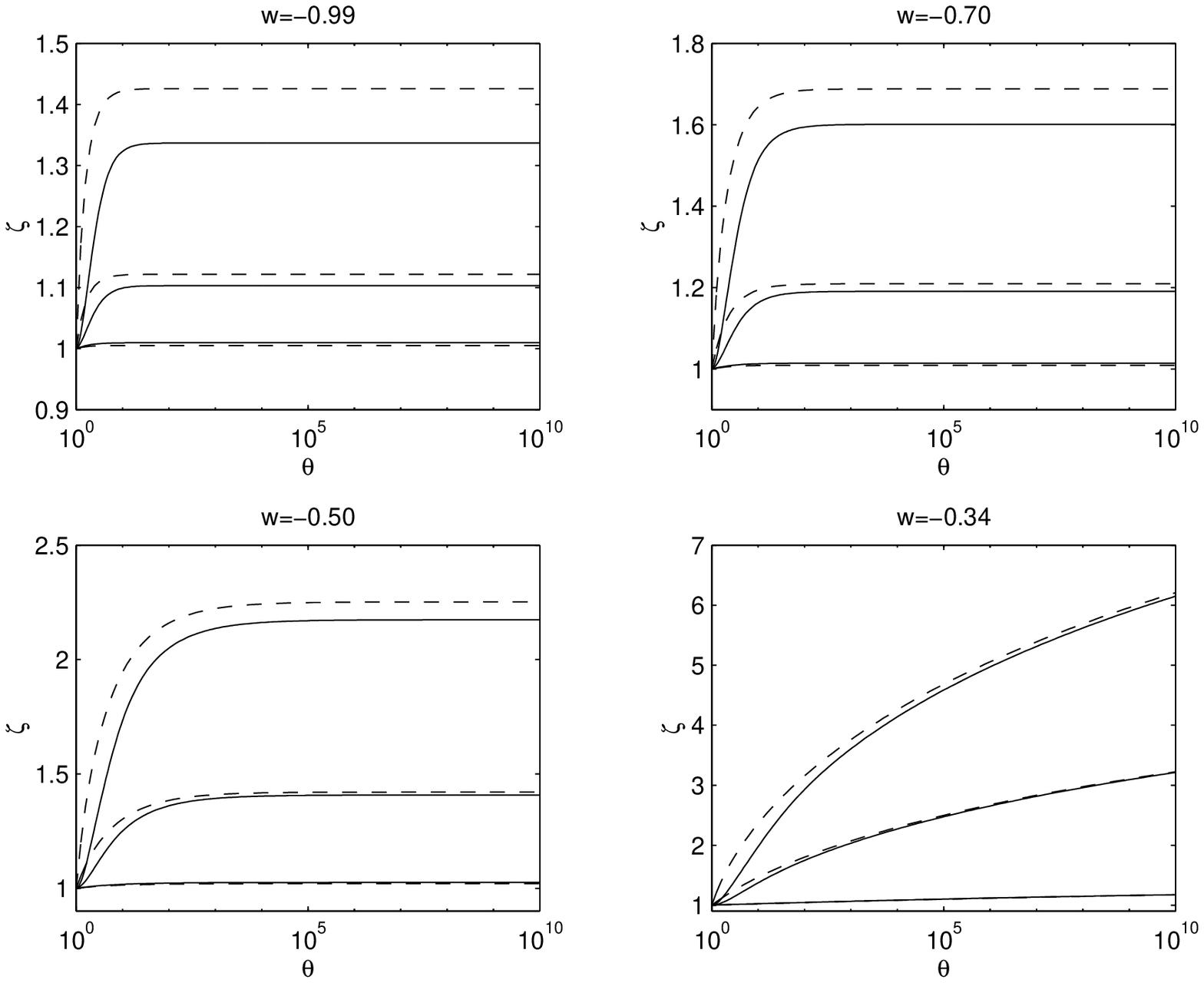}
\caption{\label{figure5}Comparing the full (numerical, see Eqns.
(\ref{vosv})) and approximate (analytic, see Eqn. (\protect\ref{sol1})) VOS model solutions for the
evolution of long strings during an inflationary epoch. 
These are respectively shown by solid and dashed lines, for each set of
initial conditions. Each panel corresponds to a different value of
the equation of state parameter $w$, and within each one the three sets
of lines correspond to an initial long-string correlation length given 
by $\epsilon=0.5, 1, 5$,
respectively from top to bottom.}
\end{figure}

We can use this model to study the evolution of the loops larger than $R_c$, and compare it with the evolution of the long-string correlation length. We solve numerically the Eqns. (\ref{vosvloop}) or (\ref{vosv}) for given values of the equation of state $w$ and the initial loop size or string correlation length, which we parametrize as
$R_iH_i=\epsilon$ and $L_iH_i=\epsilon$ respectively (note that for an initially static loop $R_i=r_i$). We expect large loops to be conformally stretched, so the relevant quantity to plot is $\zeta_{loop}=Ra_i/R_ia$ for loops of radius $R$, or $\zeta_{long}=La_i/L_ia$ for the long-string correlation length, as a function of $\theta=a/a_i$. Examples can be found in Fig. \ref{figure4} for loops, and in Fig. \ref{figure5} for the long strings; the solid lines correspond to the numerical solutions.

The difference between the two results is more apparent than real and has to do with the fact that we did not take into account an energy loss mechanism for large loops, and moreover $R$ and $L$ are not directly related to each other. Recall that $R$ is directly proportional to the loop's energy density, while $L$ is inversely proportional to (the square root of) the energy density of long strings in a given volume, $L^2=\mu V/E$. Therefore, the dynamical effect of expansion will be reflected differently in $R$ and $L$. We can see this easily by defining an effective long string energy lengthscale that is proportional to the energy, $E=\mu V/L^2\equiv \mu L_{eff}$. Substituting this definition in Eqn. (\ref{vosv}) we find
\begin{equation}
\frac{dL_{eff}}{dt}=(1-2v^2)HL_{eff}-{\tilde c}vN_{eff}\,,\label{effectivel}
\end{equation}
which is similar to Eqn. (\ref{vosvloop}) except for the loop production term (which is now negative since loop production removes energy from the long strings). We have defined $N_{eff}=L_{eff}/L$ which is the number of segments of size $L$ in the chosen volume. The energy density of both large loops and long strings will eventually scale as $a^{-2}$ instead of $a^{-3}$ (the case of non-relativistic point-like particles). Consequently both large cosmic loops and long strings have the ability to survive more e-foldings of inflation than point-like objects.

Since we know that the string motion will be damped, we can use the VOS model prediction for the relation between the velocity and the correlation length in this regime, namely \cite{quantitative,extending} $v\sim k_0/2Hr$, where $k_0=2\sqrt{2}/\pi$ is the value of the momentum parameter $k(v)$ in the non-relativistic limit $v\longrightarrow 0$. This can then be substituted in the evolution equations, and the following solutions are obtained
\be\label{sol1loop}
\zeta^2_{loop}=1-\frac{k_0^2}{\epsilon^2 (1+3w)}
\left(\theta^{1+3w}-1\right)\,,
\ee
\be\label{sol1}
\zeta^2_{long}=1+\frac{k_0(k_0+{\tilde c})}{2\epsilon^2 (1+3w)}
\left(\theta^{1+3w}-1\right)\,.
\ee
In fact these solutions are also valid for the de Sitter case, $w=-1$. The equation of state factor is negative, and hence one always has $\zeta>1$ for long strings, and $\zeta<1$ for loops. Figs. \ref{figure4} and \ref{figure5} also show (in dashed lines) this solution for comparison with the numerical solution of the full VOS model. This analytic approximation works very well for very large loops, and it becomes less accurate for relatively smaller ones. These loops do not have non-relativistic velocities throughout their evolution: they can be mildly relativistic for the first few $e$-foldings, before they suffer enough stretching to slow them down, in which case the above ansatz is a good approximation in the early stages of the evolution. On the other hand, the the loop solution (\ref{sol1loop}) also breaks down when the size is very small.

\section{\label{wallevol}Interlude: Domain walls}

We will now provide a brief discussion of the case of domain wall networks. At the conceptual level the analysis is of course entirely analogous to the case of cosmic strings, but the different co-dimension of the defects will introduce some quantitative differences, and determining where and why they emerge is helpful in the overall understanding of the problem.

Thin spherical domain walls in Minkowski space have a conserved wall invariant area given by $S_w=4 \pi \gamma r^2$ where as before $ \gamma \equiv (1-v^2)^{-1/2}$, $v$ is the domain wall velocity and $r$ is the coordinate radius of the domain wall. This implies that $R_w=\gamma^{1/2} r$ is a conserved wall invariant radius---the invariant circular cosmic string loop radius is given by $R_s=\gamma r$. Now consider the evolution of a spherical domain wall in a flat FRW universe. Its trajectory can be described by ${\bf x}(\eta,\theta) = q(\eta) \, \left(\cos \theta \cos \phi, \cos\theta \sin \phi, \sin \theta \right)$ with $0\leq \phi \leq 2 \pi$, $0\leq \theta\leq \pi$. Then the microscopic wall equation of motion can be written as
\be
\frac{dR_w}{dt}=(1-\frac{3}{2} v^2)HR_w\,,
\frac{dv}{dt}=(1-v^2)\left(\frac{g(v)}{R_w}-3Hv\right)\,;
\label{newmicvloop}
\ee
where now $g(v)=2\gamma^{1/2} sign{(-q)}$. The differences between the Hubble damping terms in the string and wall equations are related to the fact that the domain wall momentum per unit comoving \textit{area} is proportional to $a^{-1}$ in the case of an infinite planar domain wall (so that $v_w \gamma \propto a^{-3}$) while in the case of an infinite straight string it is the momentum per unit comoving \textit{length} that is proportional to $a^{-1}$ (so that $v \gamma \propto a^{-2}$).  

As for circular loops, the motion of spherical domain walls in de Sitter space is periodic as long as the domain wall is small compared with the Hubble radius, that is
\be\label{newhalfsize}
r_i H < R_c H=\frac{2}{3}\,.
\ee
Again, notice that if $H$ is a constant then Eqns.  (\ref{newmicvloop}) are invariant with respect to the the transformation $v \to -v$, $g \to - g$ which implies that if the wall is able to collapse then the wall motion will necessarily be periodic. For periodic solutions $\langle d(\ln R)/dt \rangle = 0$ and consequently $\langle v^2 \rangle = 2/3$ if $H$ is a constant. This happens because the energy density in sub-critical periodic domain walls should evolve as matter (with $\rho \propto a^{-3}$) assuming a negligible energy transfer from super to sub-critical domain walls. The equation of state parameter for a domain wall gas is $w \equiv p/\rho =-2/3 + \langle v^2 \rangle$ and consequently $\langle v^2 \rangle = 2/3$ yields $w=0$.

One can similarly find a stationary solution of equations of motion with fixed physical radius and velocity, representing a contracting domain wall standing still against the Hubble expansion. It has the form
\be
Hr = \sqrt{3/8}\,,\qquad {\dot r} = 0\, \label{newstatphys}
\ee
where $r=a q$ is the physical radius, and in comoving coordinates ${\cal H}q = \sqrt{3/8}$, $v = {\dot q} = - \sqrt{2/3}$. These are the analogues of Eqn. (\ref{statphys}). For walls with $r_i > R_c$ the motion is no longer periodic and the domain wall will soon become frozen in comoving coordinates. Specifically, the domain wall will asymptotically behave as
\be\label{newlargerr}
q=const. \Longleftrightarrow R\propto a\,,\quad
v\propto a^{-3}\,,
\ee
and again if $H$ is not constant and/or for of non-spherical domain walls the wall motion will not be periodic.
The de Sitter case is again the best possible context for domain wall survival (assuming $w \ge -1$). In all other cases, domain wall survival will be more difficult and we expect that domain walls with $R < R_c$ will decay into radiation. Finally, notice that the energy density of large domain walls will eventually scale as $a^{-1}$ instead of $a^{-3}$ (the case of non-relativistic point-like particles) or $a^{-2}$ (the case of long strings). Consequently cosmic domain walls have the ability to survive more e-foldings of inflation than point-like objects or strings.


\section{\label{conc}Conclusions}

We studied the evolution of simple cosmic string loop solutions during an inflationary era, showing that large loops and long strings can survive more $e$-foldings of inflation than point-like objects. In a de Sitter universe, circular loops below an initial invariant radius $R_c=H^{-1}/2$ (assuming $v_i=0$) have periodic motion, while larger loops freeze in comoving coordinates. Realistic small loops will simply decay into gravitational radiation while large loops are stretched by the inflationary expansion. We have also shown that loop solutions with $r_i = R_c$ will asymptote a stationary solution with constant velocity and invariant loop radius. These are very special loop solutions which we expect to be uncommon even in the context of string evolution in an inflationary era where the small scale structure can be significantly reduced. For non circular loops or varying $H$ the loop motion is no longer periodic even in the case of small loops.

Our results should provide a very good approximation to the overall behavior of realistic loops produced by a string network, and are also relevant for the final distribution of cosmic strings loops nucleated during an inflationary era which usually assumes that loops are simply stretched by the expansion. It has been shown \cite{openinf,anisot} that a cosmic string network can survive up to 60 e-foldings of inflation: even though it is pushed outside the horizon, its evolution once inflation ends ensures that it will be back inside the horizon in time to have observable consequences. Large loops produced during the final stages of inflation must therefore be included in any quantitatively realistic study of these scenarios. A detailed analysis of these issues is left for future work.

\begin{acknowledgments}
This work was done in the context of the ESF COSLAB network and funded by FCT (Portugal), in the framework of the POCI2010 program, supported by FEDER. Specific funding came from grant POCI/CTE-AST/60808/2004. C.M. is grateful to the Galileo Institute for Theoretical Physics for hospitality, and to INFN for partial support during the completion of this work.
\end{acknowledgments}
\bibliography{loopinf}

\begin{thebibliography}{9}
\expandafter\ifx\csname natexlab\endcsname\relax\def\natexlab#1{#1}\fi
\expandafter\ifx\csname bibnamefont\endcsname\relax
  \def\bibnamefont#1{#1}\fi
\expandafter\ifx\csname bibfnamefont\endcsname\relax
  \def\bibfnamefont#1{#1}\fi
\expandafter\ifx\csname citenamefont\endcsname\relax
  \def\citenamefont#1{#1}\fi
\expandafter\ifx\csname url\endcsname\relax
  \def\url#1{\texttt{#1}}\fi
\expandafter\ifx\csname urlprefix\endcsname\relax\def\urlprefix{URL }\fi
\providecommand{\bibinfo}[2]{#2}
\providecommand{\eprint}[2][]{\url{#2}}

\bibitem[{\citenamefont{Kibble}(1976)}]{Kibble}
\bibinfo{author}{\bibfnamefont{T.~W.~B.} \bibnamefont{Kibble}},
  \bibinfo{journal}{J. Phys.} \textbf{\bibinfo{volume}{A9}},
  \bibinfo{pages}{1387} (\bibinfo{year}{1976}).

\bibitem[{\citenamefont{Vilenkin and Shellard}(1994)}]{Book}
\bibinfo{author}{\bibfnamefont{A.}~\bibnamefont{Vilenkin}} \bibnamefont{and}
  \bibinfo{author}{\bibfnamefont{E.~P.~S.} \bibnamefont{Shellard}}
  (\bibinfo{year}{1994}), \bibinfo{note}{{ }Cambridge, U.K.: Cambridge
  University Press}.

\bibitem[{\citenamefont{Avelino
  et~al.}(1999{\natexlab{a}})\citenamefont{Avelino, Caldwell, and
  Martins}}]{openinf}
\bibinfo{author}{\bibfnamefont{P.~P.} \bibnamefont{Avelino}},
  \bibinfo{author}{\bibfnamefont{R.~R.} \bibnamefont{Caldwell}},
  \bibnamefont{and} \bibinfo{author}{\bibfnamefont{C.~J. A.~P.}
  \bibnamefont{Martins}}, \bibinfo{journal}{Phys. Rev.}
  \textbf{\bibinfo{volume}{D59}}, \bibinfo{pages}{123509}
  (\bibinfo{year}{1999}{\natexlab{a}}), \eprint{astro-ph/9809130}.

\bibitem[{\citenamefont{Basu and Vilenkin}(1994)}]{basu}
\bibinfo{author}{\bibfnamefont{R.}~\bibnamefont{Basu}} \bibnamefont{and}
  \bibinfo{author}{\bibfnamefont{A.}~\bibnamefont{Vilenkin}},
  \bibinfo{journal}{Phys. Rev.} \textbf{\bibinfo{volume}{D50}},
  \bibinfo{pages}{7150} (\bibinfo{year}{1994}), \eprint{gr-qc/9402040}.

\bibitem[{\citenamefont{Martins and
  Shellard}(1996{\natexlab{a}})}]{quantitative}
\bibinfo{author}{\bibfnamefont{C.~J. A.~P.} \bibnamefont{Martins}}
  \bibnamefont{and} \bibinfo{author}{\bibfnamefont{E.~P.~S.}
  \bibnamefont{Shellard}}, \bibinfo{journal}{Phys. Rev.}
  \textbf{\bibinfo{volume}{D54}}, \bibinfo{pages}{2535}
  (\bibinfo{year}{1996}{\natexlab{a}}), \eprint{hep-ph/9602271}.

\bibitem[{\citenamefont{Martins and Shellard}(2002)}]{extending}
\bibinfo{author}{\bibfnamefont{C.~J. A.~P.} \bibnamefont{Martins}}
  \bibnamefont{and} \bibinfo{author}{\bibfnamefont{E.~P.~S.}
  \bibnamefont{Shellard}}, \bibinfo{journal}{Phys. Rev.}
  \textbf{\bibinfo{volume}{D65}}, \bibinfo{pages}{043514}
  (\bibinfo{year}{2002}), \eprint{hep-ph/0003298}.

\bibitem[{\citenamefont{Avelino
  et~al.}(1999{\natexlab{b}})\citenamefont{Avelino, Shellard, Wu, and
  Allen}}]{protyloops}
\bibinfo{author}{\bibfnamefont{P.~P.} \bibnamefont{Avelino}},
  \bibinfo{author}{\bibfnamefont{E.~P.~S.} \bibnamefont{Shellard}},
  \bibinfo{author}{\bibfnamefont{J.~H.~P.} \bibnamefont{Wu}}, \bibnamefont{and}
  \bibinfo{author}{\bibfnamefont{B.}~\bibnamefont{Allen}},
  \bibinfo{journal}{Phys. Rev.} \textbf{\bibinfo{volume}{D60}},
  \bibinfo{pages}{023511} (\bibinfo{year}{1999}{\natexlab{b}}),
  \eprint{astro-ph/9810439}.

\bibitem[{\citenamefont{Avelino and Martins}(2003)}]{anisot}
\bibinfo{author}{\bibfnamefont{P.~P.} \bibnamefont{Avelino}} \bibnamefont{and}
  \bibinfo{author}{\bibfnamefont{C.~J. A.~P.} \bibnamefont{Martins}},
  \bibinfo{journal}{Phys. Rev.} \textbf{\bibinfo{volume}{D68}},
  \bibinfo{pages}{107301} (\bibinfo{year}{2003}), \eprint{astro-ph/0306294}.

\bibitem[{\citenamefont{Martins and Shellard}(1996{\natexlab{b}})}]{first}
\bibinfo{author}{\bibfnamefont{C.~J. A.~P.} \bibnamefont{Martins}}
  \bibnamefont{and} \bibinfo{author}{\bibfnamefont{E.~P.~S.}
  \bibnamefont{Shellard}}, \bibinfo{journal}{Phys. Rev.}
  \textbf{\bibinfo{volume}{D53}}, \bibinfo{pages}{575}
  (\bibinfo{year}{1996}{\natexlab{b}}), \eprint{hep-ph/9507335}.

\end{thebibliography}
\end{document}